\documentclass{iaus}
\usepackage{graphicx}
\def\ltsima{$\; \buildrel < \over \sim \;$}
\def\simlt{\lower.5ex\hbox{\ltsima}}
\def\gtsima{$\; \buildrel > \over \sim \;$}
\def\simgt{\lower.5ex\hbox{\gtsima}}
\title[Stellar Populations with ELTs] 
{Stellar Populations with ELTs}

\author[Rosemary F.G. Wyse, Gerard Gilmore]   
{Rosemary F.G. Wyse$^1$%
\and Gerard Gilmore$^2$}

\affiliation{$^1$Department of Physics and Astronomy, Johns Hopkins University, Baltimore,
 MD~21218, USA  \break email: wyse@pha.jhu.edu\\[\affilskip]
$^2$Institute of Astronomy, Madingley Rd, Cambridge CB3 0HA, UK
 \break email: gil@ast.cam.ac.uk}

\pubyear{2006}
\volume{232}  
\pagerange{nnn}
\date{}
\jname{P. Whitelock, B. Leidundgeit \& M. Dennefeld, eds.}
\begin{document}

\maketitle

\begin{abstract}

The star formation, mass assembly and chemical enrichment histories of
galaxies, and their present distributions of dark matter, remain
encoded in their stellar populations. Distinguishing the actual
distribution functions of stellar age, metallicity and kinematics at
several locations in a range of galaxies, sampling across Hubble types
and representative environments, is the information required for a
robust description of galaxy histories. Achieving this requires large
aperture, to provide the sensitivity to reach a range of environs and
Hubble types beyond the Local Group, to provide high spatial
resolution, since the fields are crowded, and preferably with optical
performance since age-sensitivity is greatest near the main-sequence
turn-off, and metallicity-sensitivity for these warm stars is greatest
in the optical.

\end{abstract}

\firstsection 
\section{Understanding galaxy formation: the context}

\subsection{The Theory}

The working paradigm for cosmological structure formation is
gravitational instability of initially low-amplitude, adiabatic,
Gaussian and near scale-invariant density fluctuations in a universe
dominated by cold dark matter (with dark energy accelerating the
expansion of the Universe at late times).  The first structures that
form are small, perhaps equal in mass to a dwarf galaxy now, and large
galaxies result from the hierarchical merging and accretion of many
small systems.  The merging history of a typical massive-galaxy dark
halo is fairly straightforward to calculate, since only gravity is
involved. However, most simulations lack the resolution to model the
smallest scales with more than a few particles, and cannot follow how
far inside a `parent' halo a merging satellite penetrates, crucial to
determine the effect on the baryonic galaxy.  The state-of-the-art
Millenium Simulation (Springel et al.~2005) has a particle mass of
$8.6 \times 10^8\, h^{-1}$M$_\odot$ and a resolution of $5\,
h^{-1}$~kpc.  The incorporation of baryonic physics -- in particular
gas dissipation, star formation and feedback -- is much more complex,
and the results much more model dependent.  Feedback from an active
nucleus is only very recently beind incorporated into the simulations
(e.g.~Croton et al.~2006), with the establishment of the correlations
between black hole properties and stellar bulge/spheroid
(e.g.~Gebhardt et al.~2000; Ferrarese \& Merritt 2000) and recognition
of their probably fundamental nature (e.g.~Silk \& Rees 1998;
Kauffmann \& Haenelt 2000).  This of course adds more uncertain
physics and associated parameters into the models.

Abadi et al.~(2003) present a recent high-resolution simulation of the
formation of a present-day disk galaxy that demonstrates many of the
important aspects, including the outstanding problem of how to include
star formation and gas physics.  Generic predictions for disk galaxies
include the following:

\begin{itemize}

\item{} Extended disks form late, after a redshift of unity, or a
lookback time of $\sim 8$~Gyr, in order to avoid losing too much
angular momentum during active merging at earlier times

\item{} A large disk galaxy should have hundreds of surviving
satellite dark haloes at the present day

\item{} The stellar halo is formed from disrupted satellites  

\item{} Minor mergers (a mass ratio of $\sim 10- 20$\% between the
satellite and the disk -- a much smaller ratio between the satellite
and the larger host dark halo) into a disk heat it, forming a thick disk
out of a pre-existing thin stellar disk, and create torques that drive
gas into the central (bulge?) regions

\item{} More significant mergers transform a disk galaxy into an S0 or
even an elliptical

\item{} Subsequent accretion of gas can reform a thin disk 

\item{} Stars can be accreted into the thin disk from suitably massive
satellites (dynamical friction must be efficient) and if to masquerade
as stars formed in the thin disk, must be on suitable high angular
momentum, prograde orbits

\end{itemize}

Elliticals form by `major mergers', with a mass ratio of approximately
unity; there is much uncertainty (freedom in the models) about the gas
fraction of the merging entities, and how much star formation and
`feedback' occurs during the merger (e.g. Larson \& Tinsley 1979;
Zurek, Quinn \& Salmon, 1988; Kauffmann 1996; Cole, Lacey, Baugh \&
Frenk 2000). 

Dwarf galaxies have the most fragile global potential wells, and are
expected to be the most strongly affected, in any model, by internal
feedback processes, by external ionization and/or ram pressure
stripping (e.g.~Sandage~1965; Saito 1979; Wyse \& Silk 1985; Dekel \&
Silk 1986; Efstathiou 1992; Bullock, Kravtsov \& Weinberg 2000;
Robertson et al.~2005).  Detailed baryonic astrophysics has been
appealed to, to solve the well-established predicted excess of
satellite halos in CDM models, compared to observed satellite galaxies
(Moore et al.~1999; Klypin et al.~1999). 

\subsection{Observational Tests, Integrated Light}

 A first step in an observational approach to understanding the
physics behind galaxy formation is to obtain and then analyse large
datasets, to identify patterns and correlations as a means to
underlying physical causal connections.  The dataset could be simple
images of galaxies, and the pattern the Hubble Sequence.  More
detailed correlations such as the various projections of the
Fundamental Plane of ellipticals (Dressler et al.~1987; Djorgovski \&
Davis 1987) and the Tully-Fisher relationship for spirals and
associated dark matter scaling relationships (e.g.~Bell \& de Jong
2001) provide more insight.  The small scatter seen in these
correlations at low redshift means that late random merging of stellar
galaxies cannot be the dominant evolutionary trend. Indeed,
correlations of the properties of the stellar populations with overall
potential well depth suggest that if mergers are indeed the dominant
mechanism by which galaxies form and evolve, the mergers must be gas
rich, and most stars form during the mergers, with the last
significant merger dominating the star formation history.

Spectroscopic surveys of faint galaxies, combined with morphological
information from high-resolution HST images, have allowed
investigations of the evolution of the Fundamental Plane with
redshift, out to $z \sim 1$, for both cluster E/SO galaxies (e.g. van
Dokkum et al.~1998) and field E/SO galaxies (e.g.~Treu et al.~2005).
The conclusions (Treu et al.~2005) are that the most massive systems
have evolved passively, consistent with the bulk of star formation at
higher redshifts, $z \sim 2$, and even for lower masses, most stars
formed at these high redshift, but a significant fraction (up to $\sim
40$~\%) of stars in these systems could have formed at lower
redshifts.  The trend is more pronounced in clusters, as expected from
hierarchical clustering, but the density dependence is weaker than
model predictions. Such `downsizing' in star formation (cf.~Cowie et
al.~1996) is not a natural prediction of hierarchical clustering
models.  The dominance of early star formation in massive systems is
consistent with the results of surveys of faint galaxies selected in
the near infra-red, which found a significant population of red,
massive galaxies at redshifts of $z \simlt 2$ (Glazebrook et al.~2005;
Cimatti et al.~2005).  That major merging does not play a major role
in determining the star formation rate of galaxies over cosmic time is
also the conclusion of a survey of galaxies at redshift $\sim 0.7$,
combining Spitzer Space Telescope mid-IR observations (to determine the
star formation rate) with morphological information from HST imaging
(Bell et al.~2005).  The robust identification of
progenitor/descendant populations, and untangling number density
evolution from luminosity evolution, requires that surveys be carried
out at low, intermediate and high redshift.

Dissipationless ('dry' in current jargon) merging of ellipticals to
form more massive ellipticals appears to occur in dense clusters (Tran
et al.~2005), and in the field (van Dokkum 2005).  The relative
motions of galaxies in clusters are large compared to the depth of the
internal potential well of most galaxies, leading to the expectation
that only the most massive systems should merge (the merging
cross-section is a steeply decreasing function of the ratio of
relative velocity to internal velocity dispersion; Makino \& Hut
1997).  `Selective' merging whereby only massive systems form more
massive systems may preserve the fundamental plane at least in its
structural projections (Boylan-Kolchin, Ma \& Quataert 2005); chemical
abundances may be a more difficult aspect.

Trends with metallicity provide strong constraints on merging, both
dissipational and dissipationless.  Kauffmann \& Charlot (1998) argue
that generically in CDM models large ellipticals form from a few large
disk galaxy progenitors (rather than a large number of small
progenitors) and predict a mass-metallicity relation for ellipticals,
provided there is significant star formation and strong feedback
during mergers (see also Nagashima et al.~2005); this last point is
the crucial one, since they found that a different star
formation/feedback prescription failed to produce a mass-metallicity
trend in agreement with the observations.  Indeed Larson \& Tinsley
(1979) demonstrated that the mass-metallicity relationship for
ellipticals could be reproduced in a merger scenario provided mergers
were gas rich, and the efficiency of star formation increased with
total mass.  Another prediction, that stars in more massive
ellipticals should have a younger mean age than stars in lower
luminosity ellipticals does not appear to be in agreement with the
papers discussed above, or with the inferred star formation histories
as a function of mass from the large Sloan Digital Sky Survey (SDSS)
spectroscopic database (Jimenez et al.~2004) which show the opposite
trend (see Fig.~1 here).  However, these spectra are for the
integrated light of the central regions only, and the analysis depends
on detailed comparisons with spectral synthesis models, necessarily
involving inherent degeneracies among age distribution, metallicity
distribution and dust extinction.  An ELT with capabilities to obtain
spatially resolved spectra plus deep colour-magnitude diagrams from
resolved stellar populations will provide much superior capabilities
to determine age and metallicity distributions.  For the nearer
galaxies, spectra of individual stars will provide even more power.

\begin{figure}[h!]
\begin{center}
\includegraphics[width=4.5in]{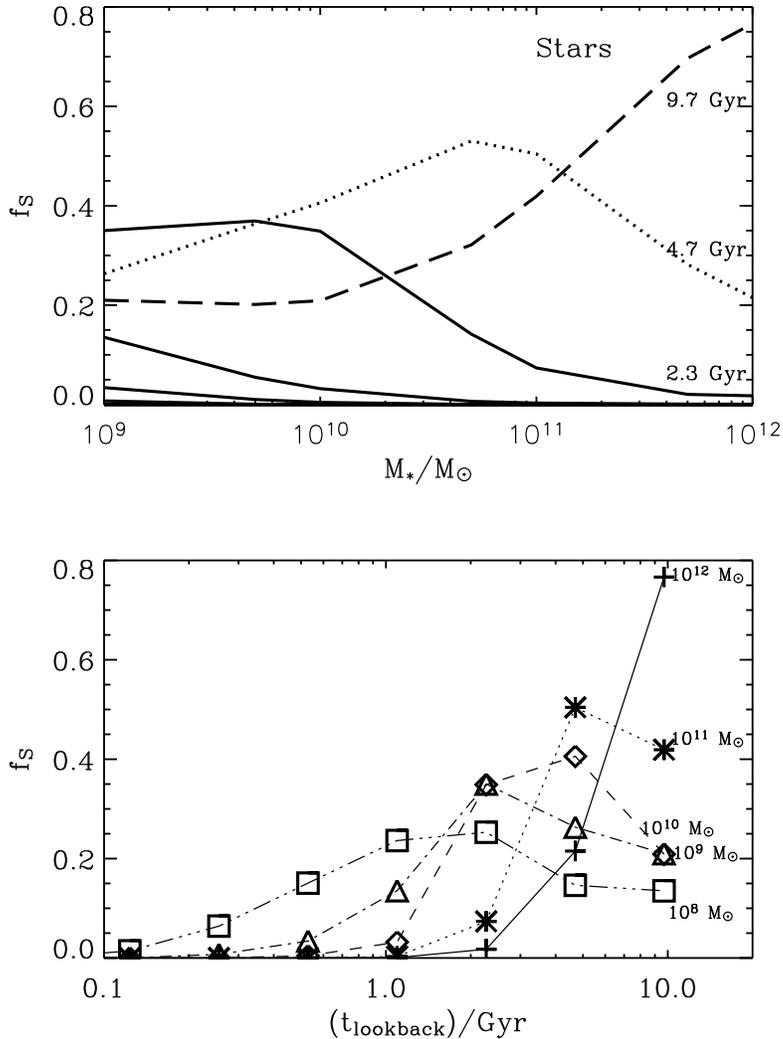}
\caption{Taken from Jiminez etal 2004, cf also Heavens et al.~2004.
Top panel: Star Formation Histories as a function of stellar mass,
plotted as fraction of stellar mass formed in each of 6 age bins, only
3 labelled.  Bottom panel: same thing but plotted as stellar mass
bins.  An ELT will enable the derivation of more detailed star
formation histories, with fewer uncertainties.}
\end{center}
\end{figure}

Kauffmann \& Charlot (1998) further predict that the mass-metallicity
relationship for ellipticals should remain `virtually unchanged out to
high redshift'.  With an ELT, we can test this prediction.

The suggestions above that the gas fraction and amount of star
formation in a merger may vary with mass, with the most massive
systems forming in dissipationless mergers, should leave signatures in
the mass-metallicity relationship.  The correlation between central
velocity dispersion and the magnesium index Mgb 
does show a hint of a flattening at the highest
values of the velocity dispersion, as expected if simple stellar
mergers with no chemical evolution occur, but there are few galaxies
to define the sample.  The early-type galaxies in the Sloan Digital
Sky Survey do not have spectra of high enough signal-to-noise for
robust measurement of line indices for each galaxy (Bernardi et
al.~2003).  Stacking of spectra of galaxies of similar velocity
dispersion (and environment) does not show evidence of a turnover
(Bernardi et al.~2003), especially when allowance is made at the
highest velocity dispersions for possible superpositions of galaxies
in the one fiber (Bernardi et al.~2005).  However, spatially resolved
spectra are required to establish the reality, or otherwise, of
interloper companion galaxies.

Tremonti et al.~(2004) studied star-forming galaxies in the SDSS
spectroscopic database, confirming a strong correlation between
present-day gas metallicity and stellar mass, with a suggestion of a
turnover for masses greater than a few times 10$^{10}$M$_\odot$.  This
is interpreted in terms of a varying importance of gas outflows rather
than anything to do with mergers.

Thus analyses of the integrated light (or, in some cases just the
central regions) together with HST-resolution structural information
of local and moderate redshift galaxies have pointed to a general
`anti-hierarchical' picture whereby large galaxies form their stars
early, and probably assembled their mass early also.  

Extended disks have been identifed by imaging in the rest-frame
optical out to redshift $\simlt 3$ (Trujillo et al.~2005), and their
sizes are consistent with little evolution since then, significantly
less than predicted by semi-analytic CDM models (Mo, Mao \& White
1998), but consistent with the simplest picture of gaseous infall and
star formation within a fixed potential well, with the star formation
rate higher in the central disk.  Indeed, the interpretation from
these high redshift observations is that `stellar disks form from
early on, in large haloes' (Trujillo et al.~2005).

The lowest mass galaxies can be studied in the Local Group, where the
metallicity-luminosity relation is extremely well-defined (see
e.g.~Dekel \& Woo 2003).  However, this is not a straightforward
metallicity--mass relation, since many of the gas-poor dwarfs
apparently have little variation in total mass, with the internal
kinematics pointing to a total (dark) mass of around $10^7$M$_\odot$
(e.g.~Mateo 1998, his Fig.~9).  This peaked mass function is not at
all expected in CDM; note that while one can change the luminosity
function significantly (see Cooray \& Cen 2005) by appealing to
various kinds of feedback, the mass function not so amenable to such
modifications.

\section{Resolved Disk Galaxies: The Four(?) Stellar Population Types}

Much more astrophysics can be derived from studies of individual
stars.  The large galaxies of the Local Group are accessible to study
with current telescopes, and indeed the first significant
spectroscopic studies of significant samples of stars in M31 and M33
are underway (e.g.~Ferguson et al.~2006 and references therein;
Reitzel, Guhathakurta \& Rich 2004), complemented by wide-area shallow
imaging (Ibata et al.~2001; Ferguson et al.~2002) and deep
narrow-field imaging with the Hubble Space Telescope (Brown et
al.~2003; Ferguson et al.~2005; Brown et al.~2006). The fascinating
results from these observations -- there is significant substructure
in all stellar components, in all their properties -- underline the
need for large samples of stars with good spectra, for metallicities
and kinematics, over as much of the face of the galaxy as possible,
with matched deep colour-magnitude data.

{\bf \it This is what we desire from an ELT, for as broad a range of Hubble
types as possible.}

Such datasets will exist in the near future for the Milky Way, as a
results of the new efforts to map its stellar content (e.g.~RAVE,
SDSS-II/SEGUE, surveys with AAOmega, WFMOS...).  We already have
sufficient knowledge to use the Milky Way as a template disk galaxy,
and identify four stellar populations with properties that constrain
the star formation history, the chemical evolution history (flows,
feedback..)  and the mass assembly history.  There may even be a fifth
type, Pop III, as yet undetected but suspected. Comparing and contrasting
external galaxies with the Milky Way then constrains these crucial
aspects of galaxy formation and evolution.

\begin{itemize}
\item{The thin disk, also known as Baade's population I.  This is
composed of stars and gas on high angular momentum orbits, moving
about the center with close to the circular velocity, and thus with
only low amplitude random motions. Such a cold thin system probably
formed by dissipational collapse of gas, in a potential that is at
most adiabatically changing, and conserved angular momentum to spin-up
as it collapsed (see Fall \& Efstathiou 1980; Mo, Mao \& White 1998).
Hierarchical merging models however predict significant angular
momentum transport and produce generically disks that are too small
(Navarro \& Steinmetz 1997).  Appeal to `feedback' can prevent much of
the angular momentum losses from the proto-disk, but at the expense of
delaying the collapse to centrifugal equilibrium (e.g. Eke, Efstathiou
\& Wright 2000) and thus predicting few old stars in disks, and no
extended high-redshift disks.  Contributions from accretion of stellar
systems into the thin disk plane is possible, and predicted in some
models (e.g.~Abadi et al.~2003).  Identification of such substructure
is complicated by the fact that dynamical perturbations certainly
exist in the form of spiral arms and giant molecular cloud complexes,
and create `moving groups' (e.g.~Famaey et al.~2005).  However,
disruption of a satellite is a viable explanation for the apparent
`ring' of stars seen in the very outer regions of the Galactic disk,
if it is real (e.g.~Newberg et al.~2002; Ibata et al.~2003).
Perturbations to the thin disk cannot be too strong however, or the
disk will be destroyed (e.g.~Ostriker 1990) and thus the properties of
stars in the thin disk constrains merging histories and other
energetic dynamical processes.  The age and metallicity distributions
of the disk, well-defined only at the solar neighbourhood, point to
extended infall of metal-poor gas, and steady star ormation from a
redshift of $\simgt 1.5$ (e.g.~Binney et al.~2000) to the present.}

\item{The thick disk - this was identified as a separate component
some 25 years ago (Gilmore \& Reid 1980).  The dominant population is
old, as old as the globular cluster 47~Tuc, $\simlt 12$~Gyr, and of
intermediate metallicity in the mean, $[Fe/H] \sim -0.6$, with a
significant spread.  The chemical enrichment history revealed by the
pattern of element ratios is distinct from that of stars in the thin
disk (Bensby et al.~2004).  A plausible origin for the thick disk is
the heating of a pre-existing thin disk by a violent dynamical event
such as a minor merger; the old mean age for the thick disk limits
such events to have occurred only long ago, an important constraint --
and a problem, if found to be a typical result -- for CDM
models. Thick disks are now observed in resolved stars in other
galaxies (e.g.~Mould 2005; Yoachim \& Dalcanton~2005) but their
properties remain to be robustly determined.  An ELT would allow such
detailed studies beyond the Local Group.}

\item{The central bulge - this too was not in the classic Baade list
of stellar populations. The dominant stellar population in the bulge
of the Milky Way is old and metal-rich, with a broad spread in
metallicities.  Elemental abundances are available for remarkably few
stars, given the capabilities of current telescopes, and point to a
fairly rapid enrichment, being dominated by products of Type II
supernovae. This, together with the old age and high (phase-space)
density, point to in situ formation, in a `starburst', at high
redshift. Could this be connected to the formation of the supermassive
black hole at the Galactic Center?  The relationships between the
`bulge', the `bar' and the inner disk remain unclear.}

\item{The stellar halo, also known as Baade's Population II.  This is
a dominantly old and metal-poor component, with Type II dominated
element ratios, indicating a short duration of star formation in each
of the star-forming entities that created the halo. The outer parts
show indications of significant accretion, most dramatically due to
the Sagittarius dwarf (Ibata, Gilmore \& Irwin 1994; Majewski et
al.~2003), which is mostly intermediate-age and more metal-rich.
Accretion to the dominant Population II halo can only have occured at
early times (Unavane, Wyse \& Gilmore 1996).  The Population II halo
may be connected to the stellar bulge; one can tie gas outflow from
halo star-forming regions, required to provide the low mean
metallicity, to gas inflow to the central regions to form the
bulge. The low angular momentum of proto-halo material means that it
will only come into centrifugal equilibrium after collapsing in radius
by a significant factor.  The predicted mass ratio of bulge to halo is
around a factor of ten, just as would be expected, and the specific
angular momentum distributions of stellar halo and bulge match (Wyse
\& Gilmore 1992; see Figure~2 here).  We have yet to obtain the data
to allow a study in detail of the bulge-stellar halo connection in
external galaxies; an ELT would allow this.}

\item{Population III -- which we take to mean stars formed from
primordial gas, most probably precursers to galaxy formation.  Where
are the low-mass Pop III stars?  On-going searches for extremely low
metallicity stars in the Galactic halo have not found any strong
indications of a separate population (e.g.~Beers et al.~2005), but
have identified a few stars with extreme deficiencies in iron, and
relatively strong carbon (e.g.~Aoki et al.~2005).  The origins of this
abundance pattern are unclear. There is little observational evidence
in favour of significant variations in the stellar IMF for any of the
components discussed above, but there is strong theoretical prejudice
that primordial stars form with a narrow range of masses, around $\sim
200$~M$_\odot$ (e.g. Bromm \& Larson 2004).  The supernovae from such
stars would provide elemental abundance patterns in the stars they
enrich that do not match those of the extremely metal-poor stars.  An
ELT could perhaps see massive Population III starbursts at high
redshift.}

\item{and the dark matter - how is this related, and is its physics
really trivially simple??}

\end{itemize}

\begin{figure}[h!]
\begin{center}
\includegraphics[width=4.5in]{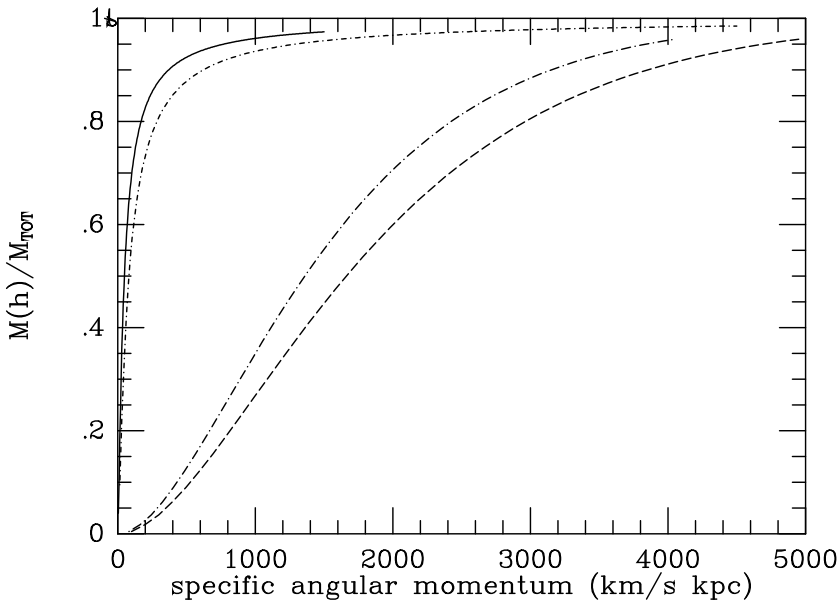}
\caption{Adapted from Wyse \& Gilmore 1992, their Figure~1.  Angular
momentum distributions of the bulge (solid curve), the stellar halo
(short-dashed/dotted curve), the thick disk (long-dashed/dotted curve)
and the thin disk (long-dashed curve).  The bulge and stellar halo
have similar distributions, as do the thick and thin disks. Does this
hold for external galaxies, pointing to fundamental relationships
between bulge and halo, and thick and thin disks? An ELT with an IFU
could tell us. }
\end{center}
\end{figure}

\section{Implications for ELT Capabilities}

It is not our purpose here to quantify a science requirements document
for any specific telescope. Rather, we end by noting that the wide
range of science questions briefly introduced above require
observations beyond the Local Group, and observations with high
sensitivity and high spatial resolution. It is a reasonable assumption
that extant 8-10m telescopes will be developed and used to their
limits, with the same set of next-generation enhancements that are
also the learning curve for the ELT-instruments and capabilities.

International technological developments, using the best 8-10m
telescopes as test-beds, will both optimise scientific gains for
today's astronomy, and allow development of next generation
facilities. The science questions introduced above are those we
believe will survive our best efforts over the next decade, and truly
need next generation capabilities.

Extending current efforts to the ELTs can push galaxy formation
understanding into contact with reality, with testable predictions,
and real science as ambitious outcomes.

\end{document}